\documentclass{article}
\usepackage{url}
\usepackage{graphicx}
\usepackage{calligra}
\usepackage[T1]{fontenc}
\usepackage{float}
\restylefloat{table}
\begin{document}
\vspace*{-3cm}
\begin{flushright}
\today

\end{flushright}


\vspace{5.0ex}

\begin{center}
  
\begin{Large}
  {\bf Raster scan or 2-D approach?}
  
\end{Large}

\end{center}

\begin{center}

\vspace{4.0ex}

\vspace{1.0ex} {\Large Louis Lyons}\\

\emph{Blackett Lab., Imperial College, London SW7 2BW, UK} \\ 
and\\
\emph{Particle Physics, Oxford OX1 3RH} \\

\vspace{4.0ex}

ABSTRACT

\end{center}
We consider the relative merits of
two different approaches to discovery or exclusion of new phenomena, a raster scan or a 2-dimensional approach.

\vspace{0.3cm}

\section{Introduction}

In this note, we consider two different types of method that can be applied in searches for new particles 
or new phenomena.
 An example would be the search for some hypothesised particle, such as the recently discovered Standard
Model (S.M.)  Higgs boson. For such a search, the data  could be a mass 
histogram or the corresponding individual mass values, but more realistically would be augmented 
by other relevant kinematic variables. Then the aim is to try to distinguish between two hypotheses: that
the data are due to some background processes ($H_0$), or that there  is also the production of some 
new particle ($H_1$)  \footnote{ Slightly confusingly, in this example the Higgs boson $H^0$ is not considered to be
part of the null hypothesis $H_0$. }. The former would result in  some relatively  smooth mass distribution, while the more exciting
production of a new particle would produce a fairly sharp peak in mass spectra.
We regard the new particle as being characterised by its mass $m_H$ and by its production
cross-section $\sigma \ $\footnote{We use the symbol $\sigma$ both for cross-section and for standard deviation. 
The meaning should be abundantly clear from the context. }. We use this example through most of this note, but a 
couple of other situations involving new physics are discussed in Section 2.

We consider two possible approaches:
\begin{itemize}
\item{ Raster scan.  Here we search over the physically interesting range of masses, and at each one separately
we make a decision as to whether we can claim a discovery (i.e. decide that $H_0$ is excluded, and 
that $H_1$ is acceptable); exclude the existence
of the new particle (reject  $H_1$); or be unable to discriminate between them. This is done separately at each possible
mass in the range, before moving on to the next mass -  see Section 3.1.}
\item{ 2-D approach. A parameter-determination method is used for $m_H$ and $\sigma$, and the choice between 
$H_0$ and $H_1$ is based on the selected region in ($\sigma,m_H$) space - see Section 3.2. }
\end{itemize}
It is to be noted that these two approaches use essentially the same data statistic $t$ (e.g. a likelihood ratio). The
difference is that the 2-D approach looks for a 2-D region around the optimal values of $\sigma$ and $m_H$, 
while the raster scan determines a best region of $\sigma$ at each $m_H$ separately.

Our aim is to compare the relative merits of these two approaches.

\section{Examples of Actual Searches}
Here we mention briefly 3 examples of searches for new physics. Throughout this note, we use the language 
relevant for a Higgs search, but the `translation' to our other two examples can be achieved via Table 1.
These are not intended to be exact descriptions of the procedures used in actual searches, but more generic
principles of approaches. For example, real searches for the Higgs boson involve several possible different
decay modes; more data variables than just the mass $M$ for each event; uncertainties on energy scales, 
backgrounds and other systematics; etc.
\begin{itemize}
\item{Higgs search.

This was performed using a raster scan \cite{CMS_Higgs, ATLAS_Higgs}. At each closely spaced mass 
over the relevant range, a separate search was performed. Fig. 1 shows the $p$-values for the null 
hypothesis as a function of $m_H$. Based on this (and on a comparison of $\sigma$ with the Standard Model 
prediction, as well as branching ratios for various decay modes), the discovery of a Higgs particle with mass 
around 126 GeV was claimed.}

\item{$B_s$ oscillations\cite{CDF_Bs, D0_Bs}.

These are parametrised by the amplitude $A$ of oscillations and their frequency, which is proportional
to the mass difference $\Delta m_s$  between the  $B_s$ and its anti-particle. A raster 
scan was performed to determine $A$ at each mass difference $\Delta m_s$ - see fig. 2. If such oscillations occur,
the value of $A$ should be unity at the relevant $\Delta m_s$, whereas otherwise $A$ is expected to be 
close to zero. On the basis of the data,
the discovery of $B_s$ oscillations at a mass difference of $\sim 17 ps^{-1}$ was claimed.}

\item{Neutrino oscillations\footnote{In contrast to the other two types of searches, plots of the accepted regions in parameter space for neutrino
oscillations tend to have the (logarithm of the) mass parameter plotted on the $y$-axis.}.

In a simplified model with only two species of mixing neutrinos, their oscillations are specified by an 
amplitude $\sin^2(2\theta)$ and by their frequency which is proportional to 
$\Delta m^2$, the difference in mass squared of the two neutrino types. A given set of data can be analysed
by obtaining the best value of $\sin^2(2\theta)$ at each value of $\Delta m^2$ (i.e. a raster scan), and 
seeing whether sin$^2(2\theta) \,= \, 0$ is excluded. 
However, Feldman and Cousins\cite{F_C} point out that for extracting  
$(\sin^2 (2\theta), \Delta m^2)$  a 2-D parameter determination is preferable (see Fig 3).
 }

\end{itemize} 

\begin{table}
\caption{Examples of searches for new physics, characterised by a strength $\phi_1$ and a mass parameter $\phi_2$.}
\vspace{0.3cm}
\centering
\begin{tabular}{|c| c| c | c| }
\hline
Data and params & Event variables & Strength $\phi_1$ &  Mass param $\phi_2$ \\
Example &  &  &  \\
\hline & & & \\
Higgs   & Mass $M$ &  Cross-section $\sigma$   &   Mass $m_H$  \\  \hline & & & \\
$B_s$ oscillations & Length, energy, decay mode  & Amplitude $A$  &  Mass diff $\Delta m_s$  \\   \hline  & & &  \\
$\nu$ oscillations  &  Length, energy, interaction type  &Amplitude $\sin^2(2\theta)$  &  Diff in mass-sq $\Delta m^2$ \\
\hline
\end{tabular}
\end{table}

An interesting question is why  the search for $B_s$ oscillations uses a raster scan, but for neutrino oscillations a 2-D approach is used.
The main reason is that the emphasis of the analysis in the neutrino oscillation case was to determine the values of the parameters
$\sin^2(2\theta)$ and $\Delta m^2$, while in the $B_s$ case it was whether the $B_s$ oscillations could be discovered. Also 
the amplitude in the $B_s$ case is expected to be unity, while for the neutrinos any value from zero to unity is physically possible. 

Another interesting contast between searches for these two types of oscillations is the difference in the 
appearence of the raster scan in the two cases. At small mass differences, the raster scan for $B_s$ gives values 
of the amplitude close to zero. In contrast, the neutrino oscillation amplitude in the corresponding region can be  large. The 
reason is that the range of observable times in the $B_s$ case covers several complete oscillations, while for some 
neutrino experiments,  only a small part of the first oscillation can be observed\footnote{In other neutrino experiments, 
the distance and resolution are such that only a time-averaged oscillation rate is observed, giving rise to a widish range of 
possible values for the mass parameter.}. 
Thus the $B_s$ data tend to pick out the correct oscillation frequency, with the fitted
amplitude at smaller frequencies being close to zero; while neutrino experiments in this regime are sensitive essentially only to the
product $\ sin^2(2\theta)\times (\Delta m^2)^2$, so smaller $\Delta m^2$ values need a larger amplitude.


\section{Further Details} 
There are several possible choices for the way a preferred region in parameter space is defined, whether this 
is for just $\sigma$ for the raster scan, or for both $m_H$ and $\sigma$ for the 2-D approach. A partial 
list of possible methods is:
\begin{itemize}
\item{Maximum likelihood approach, with the selected region being those parameter values for which 
\begin{equation}
\Delta ln{\calligra L}\  =  \ ln {\calligra L}_{max} -  ln {\calligra L}(\sigma , m_H)  \le C, 
\end{equation}
where  $ln {\calligra L}(\sigma , m_H)$ is the 
logarithm of the likelihood for parameter values $(\sigma , m_H)$;  ${\calligra L}_{max}$ is the maximum 
of the likelihood as a function of $\sigma$ for the Raster Scan, or of both $\sigma$ and $m_H$ for the 2-D approach;
and $C$ is a constant whose value depends on
the desired nominal coverage level and on the number of free parameters. For example, for a $95 \%$ region
in a 1 or 2 parameter problem, $C$ would be 1.9 or 3.0 respectively.  }
\item{An alternative
but related approach would be to define a $\chi^2$ region such that $\chi^2 (\sigma , m_H) - \chi^2_{best}$
is below some critical value.}
\item{Frequentist approach.  This uses the Neyman construction, with parameter(s) $\mu$ as $\sigma$ for
the raster scan, or ($\sigma, m_H$) for the 2-D approach. For each value of  $\mu$,
the probability density
for observing possible data $d$ is used in conjunction with an ordering rule to construct the confidence band 
containing the likely range of data values
for the given $\mu$,
at the chosen confidence level. The  observed data  
are then used to determine the range of parameter values for which the data is likely.

With just one parameter, the ordering rule can be chosen to provide  upper or lower limits, or central regions (e.g. equal tail probabilities).

In the Feldman-Cousins version of
the frequentist approach, a likelihood-ratio is used to define the ordering  rule. This extends naturally to situations 
with more than one parameter. The Feldman-Cousins approach is called `unified' as, depending on the data, 
the pre-defined ordering rule
determines whether the 1-D preferred region will be single-sided (i.e. an upper or a lower limit), or double-sided.

Another feature of the Feldman-Cousins method is that, in contrast to the frequentist approach with 
other ordering rules, empty intervals or those consisting just of the point $\sigma = 0$ are far less likely.} 

\item{Bayesian methods. Here the likelihood function is multiplied by the prior probability density $\pi({\sigma , m_H})$ to obtain the
posterior probability density $p({\sigma , m_H})$. From this, a region can be selected in parameter space that contains the required
level $f_{Bayes}$ of integrated posterior probability density. As always the Bayesian approach requires the specification of the 
(possibly multi-dimensional) prior, and it is advisable to perform a sensitivity analysis to see how much the result depends on the 
choice of prior.

If there is only one parameter, even for  a given $f_{Bayes}$, the region could be chosen for an upper limit, lower limit, central interval
with equal probability in each tail, shortest interval\footnote{A problem with the shortest interval is that it is not independent of 
changing to a different function of the parameter
e.g. In the neutrino oscillation problem, choosing $\Delta m^2$ or $ln(\Delta m^2)$; or $\theta$ or 
$\sin^2(2\theta)$ or $\tan\theta$.}, etc. 

With more than one parameter, the obvious choice is to use a cut on the value of the posterior probability density, which in one dimension corresponds to the 
 minimum length interval. Again this produces regions which are not invariant with respect to 
reparametrisations of $(\sigma , m_H)$. There is no such problem for the likelihood, $\chi^2$ or
 Feldman-Cousins approaches.}
\end{itemize}

Frequentist methods have the advantage over other approaches that coverage is guaranteed i.e. 
the procedure is such that for a series of repeated analyses each with new data,
in the absence of experimental biases, the fraction of times that the true value of the parameter(s) will be 
contained within the preferred region will be at least that specified by the chosen confidence level.

In this note, we use the generic name `preferred region' for the 1-D or 2-D parameter regions produced by any of 
these methods. Clearly, any test to see whether $\sigma$ is consistent with zero cannot employ a 
lower or an upper limit technique; equi-tailed central intervals may be problematic too.



\subsection{Raster Scan}
\subsubsection{Procedure}
The characteristic of a raster scan is that it is performed at each mass $m_H$ separately, and independently of 
what happens at any other mass. At each such mass, a decision is made 
to exclude $H_1$, to reject the null hypothesis $H_0$ or to make no choice. Thus at each mass $m_H$, a region of 
$\sigma$ is determined at the pre-defined confidence level, even though the overall fit to 
the data using these parameters may be much 
worse than fits at other masses; the raster scan 
ignores the fact that there may be a discrepancy, and still provides a range for $\sigma$ at each mass.
It then  checks whether these regions contain 
zero (as a test of the null hypothesis $H_0$) or/and $\sigma_{SM}$, the predicted cross-section assuming 
S.M. Higgs.
If the region excludes $\sigma = 0$, then $H_0$ is excluded at the relevant confidence level,
while if it does not contain $\sigma_{SM}$, the predicted cross-section, then $H_1$ is excluded.
It is thus  possible to claim a discovery of the Higgs at a specific mass; to 
claim discoveries at more than one mass\footnote{This demonstrates the slightly slippery approach
used by Particle Physicists in the definition of the New Physics involved in $H_1$. We start off by looking 
for the S.M. Higgs, but would not be unhappy if we discovered more than the expected one such object.
Similarly if the observed production rate was very inconsistent with zero, but not really in agreement with 
$\sigma_{SM}$, many (or most?) Particle Physicists would claim this as a discovery of a S.M.-like Higgs, 
but with a production rate different from the predicted rate according to the S.M.}; 
to exclude the S.M. Higgs over a range of masses; to 
have insufficient evidence for choosing between $H_0$ and $H_1$; or a combination of 
these.

An analogous procedure is used in the search for $B_s$ oscillations, where the amplitude $A$ is determined at each $\Delta m_s$.

Of course, a lot of detail has been ignored in the above brief description. 
Further discussion of hypothesis testing techniques, including the so-called $CL_s$  method\cite{Read}
for excluding $H_1$, can be found in ref. \cite{H0vH1}.

\subsubsection{Features}
\begin{itemize}
\item{We have already mentioned that because the raster scan treats each mass separately, the conclusion
as to what the data tell us at one mass is completely unaffected by what is happening at other masses (Contrast 
the 2-D approach below.)}
\item{Another consequence of dealing with each mass separately is that it may be possible to claim 
discoveries of more than one `S.M.' Higgs.} 
\item{The raster scan will provide a cross-section range for each mass, regardless of 
the quality of the fit. }
\item{The prefered  region defined by the raster scan for $\sigma$ at each possible $m_H$ is different 
from what would have been obtaind by determining the $m_H$ range at each $\sigma$. In the case of 
the Higgs (and $B_s$)  searches, the former is used as it  is more physically motivated, and it works better in practice; this 
might not be the case in different sorts of analyses where the two parameters might have a more symmetric
status.} 
\end{itemize}

\subsection{2-D Parameter Determination}
\subsubsection{Procedure}
As a result of a 2-D search for the preferred  region at some level (see Fig 4, where some 
possibilities are shown), some rules have to be formulated for deciding whether we can claim a  discovery;
exclude/disfavour the existence at some or perhaps even all relevant masses; or be unable to distinguish
between these possibilities.

Even apart from the choice of confidence levels, there is some degree of arbitrariness in how the 
procedure is defined. We consider the following as an example of a possible set of rules:
\begin{itemize}
\item{Discovery: We require the 2-D preferred region at the 5$\sigma$ level to exclude $\sigma = 0$; 
and also require the predicted $\sigma_{SM}$ to be within the 95\% contour. This ensures that the 
observed signal strength is as expected; a consequence of this is that an apparent signal which is 
significantly smaller or larger than predicted does not count as a discovery.

It may also be reasonable to impose a requirement that the width in $m_H$ of the discovery region
be consistent with expectation. e.g. in a channel with good mass resolution such as $H \rightarrow \gamma\gamma$ or 
$H \rightarrow ZZ \rightarrow 4$ charged leptons,  
 an apparent signal over a very wide mass region may be more
likely to correspond to a mis-modeled background than to a fundamental discovery. (This applies also to raster scans.)}

\item{Exclusion: Masses at which the 95\% region does not include $m_H$ are excluded. This includes the case where 
the single S.M.  Higgs boson has been discovered at some other mass. }

\item{No decision: This applies when the 95\% region includes both $\sigma=0$ and $\sigma =
\sigma_{SM}$}, unless the Higgs boson has been discovered at some other mass.
\end{itemize} 

\subsubsection{Features}
\begin{itemize}
\item{Usually when we quote the mass uncertainty on a new particle state, this is interpreted as a 1$\sigma$
error region. In the 2-D approach, the discovery region is defined by a 5$\sigma$ contour, which will 
usually correspond to a wider mass interval. This is not a big problem as, once the existence of the
particle is established, its mass could be determined in a separate analysis using any desired 
criterion for defining the mass uncertainty.}

\item{Because the model used for analysing the data assumes that there is only one S. M. Higgs boson, a signal
at one mass ($m_1$) affects what happens at other masses\footnote
{It is possible that the 5$\sigma$ discovery region could consist of two or even more separate 
regions, with more than one of these being consistent with Higgs discovery at that mass. Within the S. M., however,
this corresponds to an ambiguity in the location of the single Higgs, rather than to the existence of more than
one Higgs.}. Thus if there is another mass $m_2$ which in 
a raster scan also shows a 5$\sigma$ signal, $m_2$ could appear in the exclusion region in the 2-D approach
if the signal at $m_1$ has larger significance. Similarly, because of discovery of the one and only Higgs boson at
$m_1$, other masses are excluded; this includes mass regions where the experimental data has only 
very weak (or even no) discrimination to the presence or absence of a Higgs boson.

Thus  the 2-D approach with the assumption of just one Higgs boson has a higher probability than the
raster scan has of excluding the Higgs at other masses. However, in the absence of a significant excess 
throughout the mass range, the exclusion power of the raster scan is better, largely because of the higher cut on 
 the log-likelihood ratio needed in the 2-D scenario\cite{daniel}.}

\item{Similarly for discovery at the favoured mass, the 2-D $5\sigma$ region will include a wider range of cross-section
(and hence be less sensitive to discovery) than the corresponding region in a raster scan. For a frequentist method such as
Feldman-Cousins, the $p$-values for excluding zero cross-section will be global and local (for the 2-D and raster scan 
approaches respectively). }

\item{There are other situations in which the above procedure is not logical. For example, the preferred region might 
correspond to a low or a high signal rate, such that it does not qualify as a discovery claim, and yet we
still exclude all other masses outside the preferred region, even including those where we have little or no 
sensitivity. }
\end{itemize} 

\subsection{Absence of predicted signal strength}
 In some searches for the New Physics embodied in the alternative hypothesis $H_1$, there might not be a 
prediction for the signal strength. For example, the strength of gravitational waves observed on Earth depends
on various parameters of the source, including its distance from us. Similarly with neutrino oscillations, the 
strength factor $\sin^2(2\theta)$ can take on any value between zero and 1. In such situations, 
the procedures for exclusion of $H_1$ (and to some extent, its discovery too) are modified. 

\subsubsection{Raster Scan}
Because we deal with each mass separately, we can make a discovery by finding that the strength factor is 
inconsistent with zero. (Indeed it may be that this happens at more than one mass.) However, no check is possible 
that the strength is consistent with expectation, as the latter does not exist.

In contrast, there can be no excluded mass region, basically because the phenomenon could have a very 
weak strength, below the sensitivity of the experiment. It is possible, however, to produce an upper 
limit on the possible strength at each mass, which can then be built up into an exclusion region in 
(strength, mass) parameter space.

\subsubsection{2-D approach}
After the preferred region is found in 2-D parameter space, we can make the usual discovery claims, except 
that as in the raster scan, there is no possible check that the strength of the observed effect is as expected.

However, we now can have excluded mass regions, being those that lie outside  the preferred region.  This is 
possible even without a  predicted signal strength, provided $H_1$ involves the assumption that
there is only one true set of values for the parameters.

\section{Conclusions}
Table 2 contains an overview of  various features of  the raster scan and $\!$ 2-D approaches to discovery
and exclusion.  A few comments about some of the entries are worth emphasising.

\begin{table} 
\caption{Relative merits of Raster Scan and 2-D approach}
\vspace{0.3cm}
\centering
\begin{tabular}{|p{5cm}| p{5.4cm}| p{4.5cm}| }
\hline
$\; \; \; \; \; \; \; \; \; \; \; \; \; \; \; \; \; \; \; \; \; \; $  Method   & Raster Scan & 2-D \\
Features   &     &  \\
\hline
\hline
Aim  &  Determines $\sigma$ region at each $m_H$ & Determines $(\sigma,m_H)$ region    \\  \hline
Discovery criterion   & $\sigma$ region excludes 0   &   2-D region excludes $\sigma = 0$    \\  \hline
Multiple discoveries possible?   &  Yes      &    No   \\  \hline
Discovery at $m_1$ affects other $m$?   &     No     &  Yes \\  \hline
Exclusion criterion    &  $p$ or $CL_s$ <5\%  &    Region excludes $\sigma = \sigma_{SM}$  \\  \hline
Exclusion at $m_1$ affects other $m$?   &    No     &     No    \\  \hline
Possible to exclude at $m$ with no sensitivity?   &    No    &   Yes    \\  \hline
Confidence levels &  Separate for discovery, exclusion. Measure $m$ separately  &     Different for discovery, exclusion, measurement of $m$  \\  \hline
Local or global $p$?   &  Local. Needs LEE for global   &   Global   \\  \hline
Good points    & Treats each mass separately.\ \ \ \ \  \ \ \ \ \ \  \ \ \   Better sensitivity    &   Regions with poor Goodness of Fit not accepted  \\  \hline
Bad points    &  Determines $\sigma$ at $m$ where fit is poor.  Raster in $m$ different from in $\sigma$  &   Can exclude where no sensitivity  \\    \hline
Method best for...  &    Hypothesis testing    &    Parameter determination   \\ \hline
\end{tabular}
\end{table}

The fundamental difference between the two methods is that the 2-D approach assumes that there 
is (at most) one S.M. Higgs boson, while the raster scan is more flexible in that it treats each mass 
separately. This then can result in the 2-D method excluding Higgs production at masses for which
there is essentially no sensitivity to Higgs production, simply because a discovery claim is made 
at another mass. Another consequence is that part of the preferred region in $(\sigma,m_H)$ space 
from the raster scan can correspond to very poor fits of the theory to the data; in the 2-D approach,
only the best fit region is obtained.

Both methods use separate confidence levels for searches and for exclusion. When a discovery
is claimed at the 5$\sigma$ level, the 2-D approach provides a mass region over which this occurs. 
However a mass measurement should be performed separately as this traditionally uses a 68\% 
confidence level for the uncertainty. In the raster scan, the mass measurement is clearly a separate 
exercise from the Hypothesis Testing of discovery.

To associate a $p$-value with a discovery claim, we need to determine the (hopefully very low) 
probability of a background fluctuation giving rise to an effect at least as large as the one observed.
 If a frequentist approach is used to determine the preferred regions, the $p$-value for a discovery claim is 
directly determined by the method, as coverage is guaranteed; otherwise a brute-force Monte Carlo 
approach may be needed. 

There are (at least) two $p$-values
to consider: the local one, which is the probability of such a fluctuation at the mass observed for the data
signal, and a global $p$-value which incorporates the Look Elsewhere Effect (LEE)\cite{LEE} i.e. the 
chance of having such a fluctuation not only at the observed mass, but anywhere in the analysis.  
The 2-D approach most naturally lends itself to calculating the global $p$-value over the range of masses
relevant for the analysis, while in the raster scan the local $p$-value is more easily calculated, and hence needs to 
be corrected for the LEE.

%
\vspace{0.07in}
In conclusion, it appears that the raster scan provides a more natural approach to discovery or 
exclusion of New Physics, while the 2-D approach is preferable for parameter determination.

\vspace{0.15in} 
I would like to thank Bob Cousins, Jon Hays, Tom Junk, Richard Lockhart, Yoshi Uchida, Nick Wardle, Daniel Whiteson 
and members of the CMS Statistics Committee for many useful discussions and patient explanations.



\begin{figure}[p]
\begin{center}
\includegraphics[width=\textwidth]{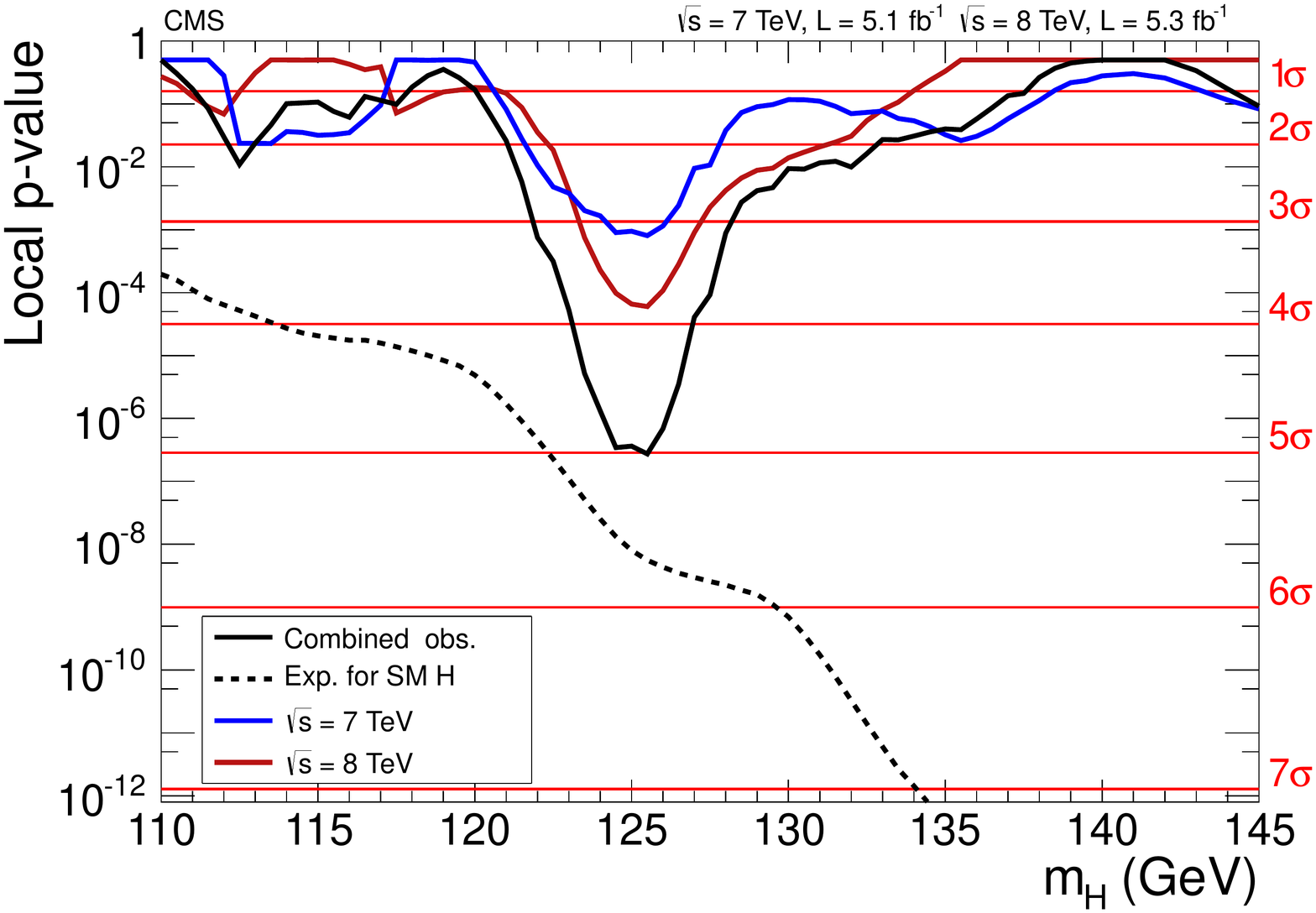}
\caption{Search for the Higgs boson. The plot from ref. \cite{CMS_Higgs}
shows the local $p$-value for the null hypothesis (`No Higgs in data, but just Standard
Model background') as a function of the supposed mass of the Higgs $m_H$. The three solid curves show the $p$-values for the 
7 TeV data, for the 8 TeV data available at the time, and for their combination. The dashed curve is the expected $p$-value 
for the combined data, assuming the Higgs exists and is produced and decays as predicted by the S. M., 
also as a function of $m_H$. The fact that in the vicinity of 125 GeV the combined $p$-value reaches 
a value below $10^{-6}$, equivalent to a $\sim5\sigma$ tail of a Gaussian distribution, and is in satisfactory agreement with 
the expected $p$-value, is taken as evidence of the discovery of a particle, consistent with the expected properties of a
Higgs boson. 
\label{fig:CMS_Higgs}}
\end{center}
\end{figure}


\begin{figure}[p]
\begin{center}
\includegraphics[width=\textwidth]{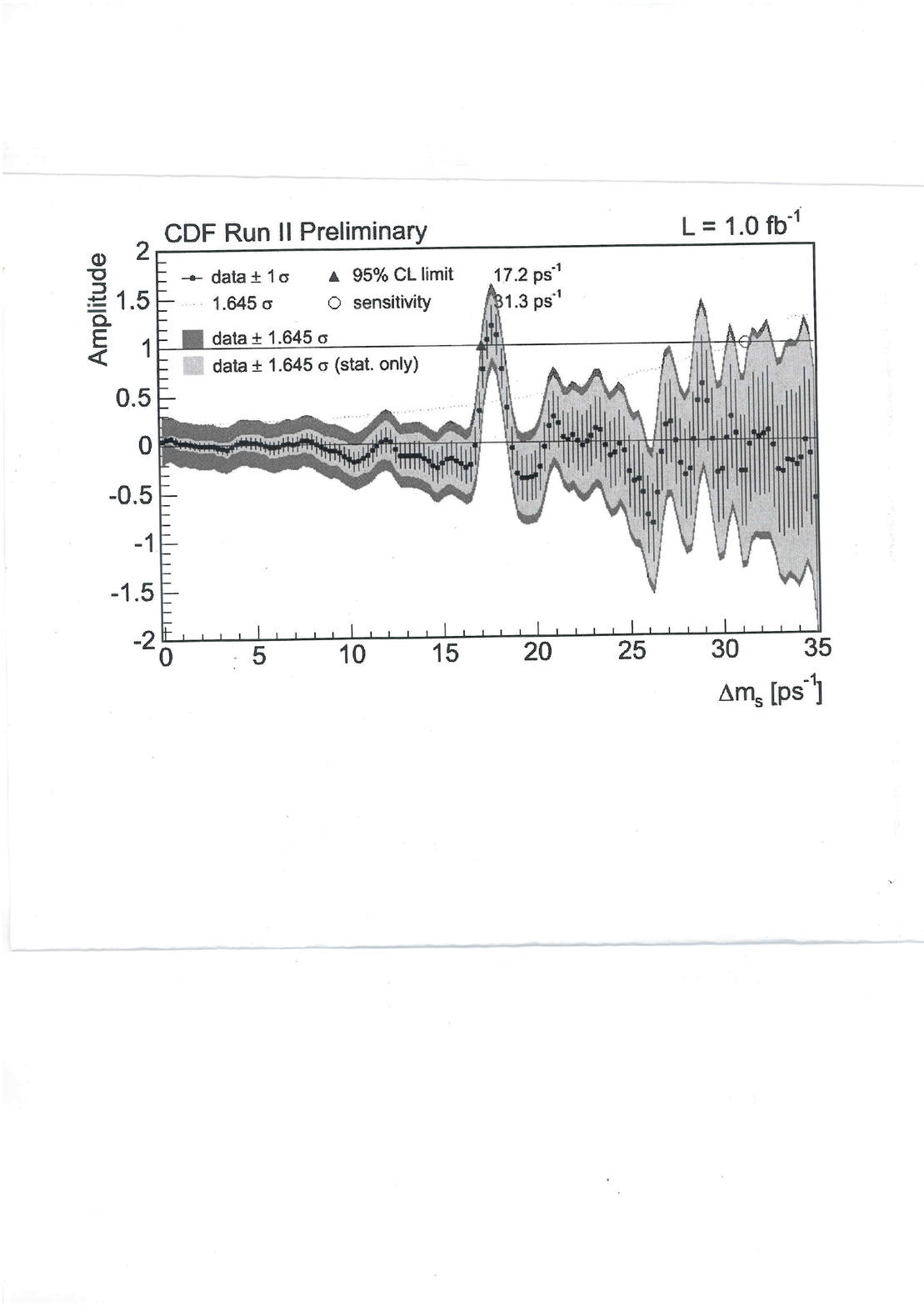}
\vspace{-2in}
\caption{Fit of $B_s$ oscillations to data from the CDF experiment\cite{CDF_Bs} at Fermilab's Tevatron. A raster 
scan is performed at each $\Delta m_s$, the mass difference between the $B_s$ and its antiparticle,
to extract the amplitude $A$ of the oscillations at that frequency. The evidence for oscillations is that for
$\Delta m_s$ near 17 $ps^{-1}$, $A$ is very far from zero, but is consistent with the oscillations'
expected value of unity.
\label{fig:CDF_Bs}}
\end{center}
\end{figure}


\begin{figure}[p]
\begin{center}
\includegraphics[width=\textwidth]{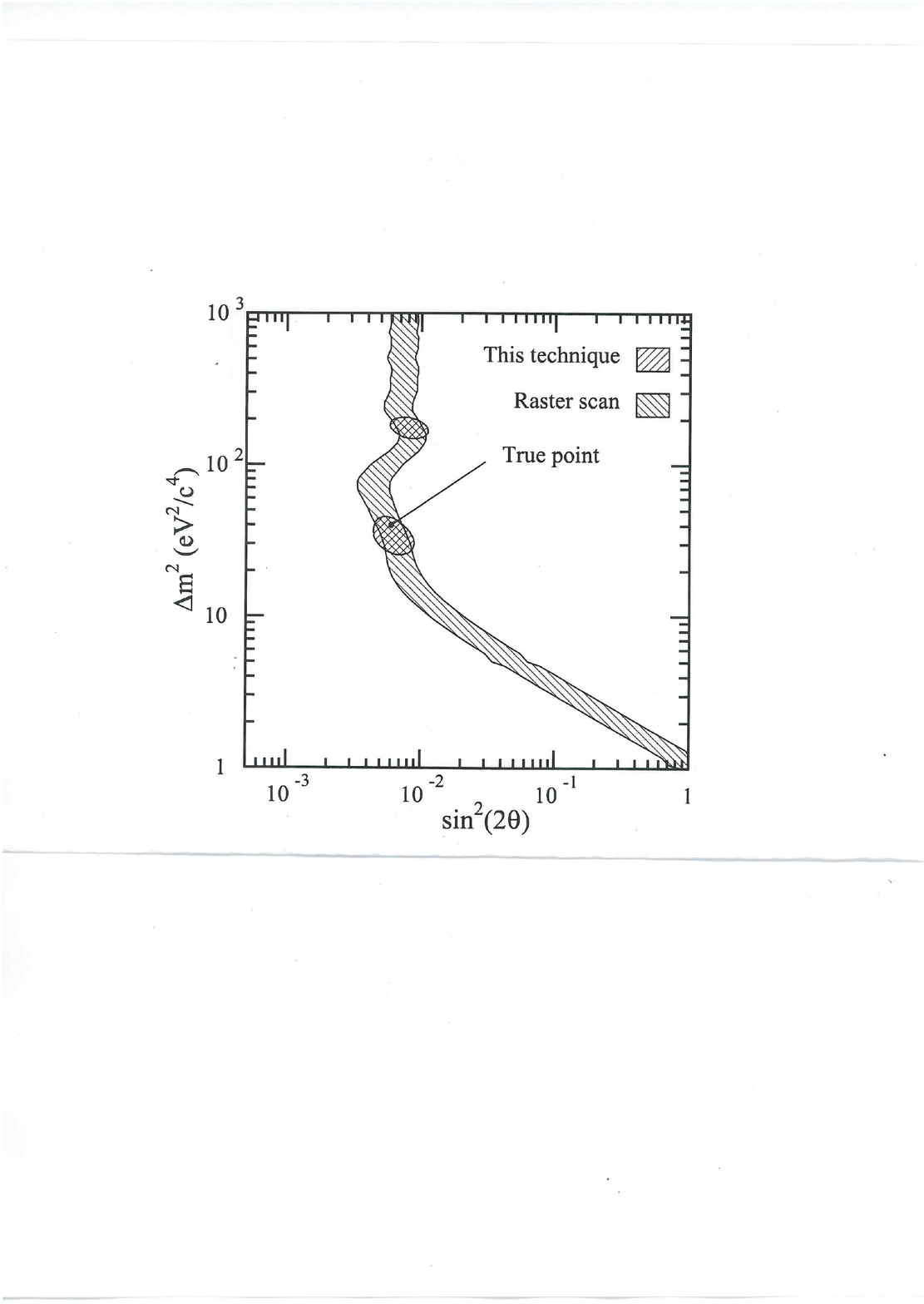}
\vspace{-2in}
\caption{ Search for neutrino oscilllations, parametrised by $\sin^2(2\theta)$ and $\Delta m^2$. The simulated  `data' were
generated assuming the parameter values denoted by `True point'. The single hatched band is the result of a raster scan
in which the best fit region in $\sin^2(2\theta)$ is determined for each value of $\Delta m^2$. The double hatched region
is the result of determining the parameters in a 2-D fit, using the Feldman-Cousins approach\cite{F_C}; the plot is taken 
from their paper. 
\label{fig:F_C}}
\end{center}
\end{figure}


\begin{figure}[p]
\begin{center}
\vspace{-2.0in}
\includegraphics[width=\textwidth]{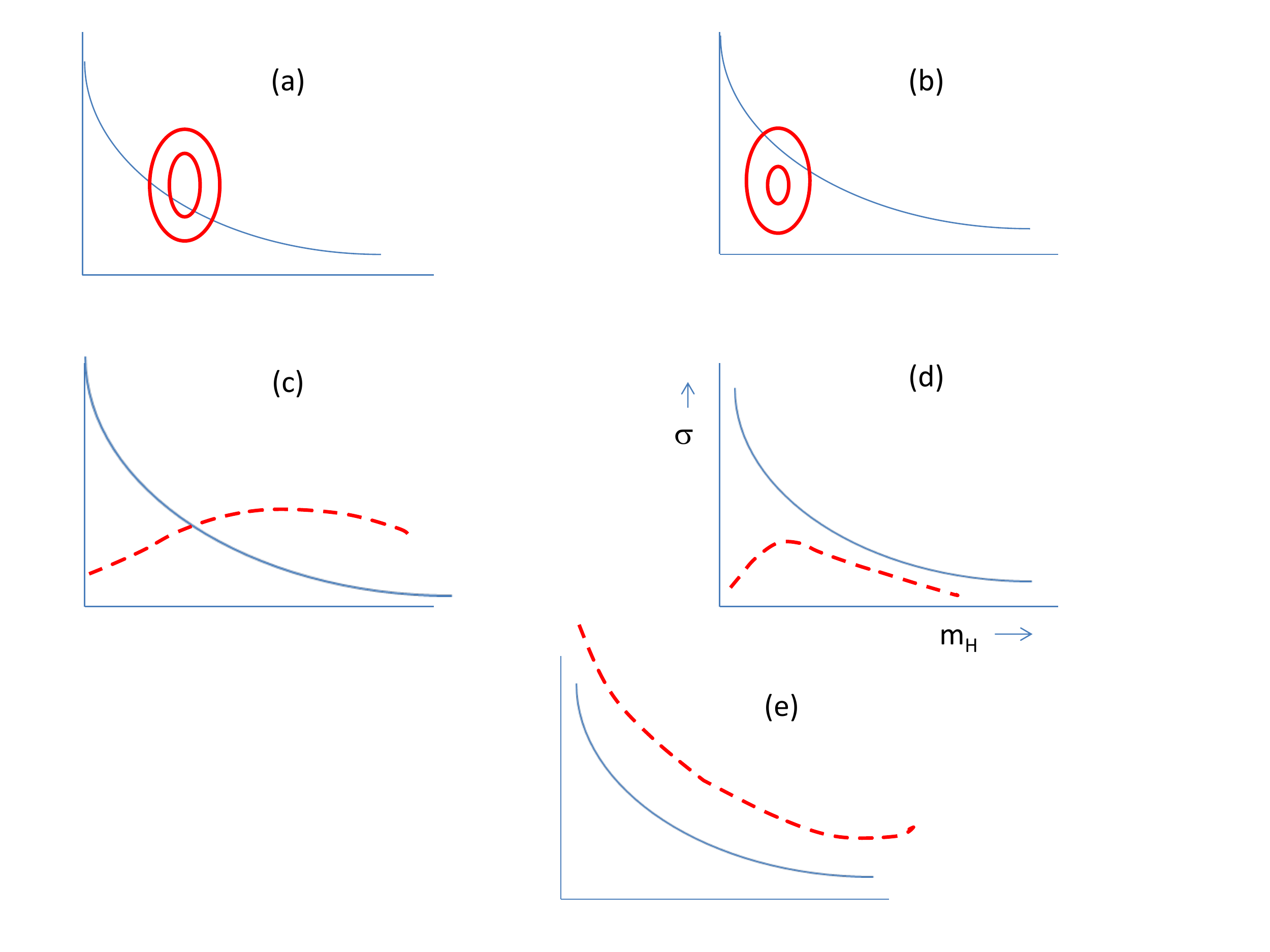}
\vspace{0.5in}
\caption{Some possible preferred regions in parameter space in a method of determining the two physics 
parameters $\sigma$ and $m_H$ simultaneously. The
solid curve is the predicted cross-section $\sigma$ at each mass, according to the Standard Model.
 The $\sigma = 0$ axis is a contour line at some acceptance level, since with zero production rate, the mass
is irrelevant. 
(a) The smaller ellipse is a 95\% CL region, and the larger one corresponds to 5$\sigma$. Since the larger ellipse 
does not reach down to the $\sigma = 0$ axis,  this corresponds to a discovery claim, with
masses outside the smaller ellipse being excluded.
(b)  Although the 5$\sigma$ ellipse excludes a zero cross-section, the 95\% CL one does not include the predicted 
cross-section, and so the hypothesised particle is excluded at all masses. This could be due to the production of a new particle,
but at a lower rate than predicted by theory.
(c) The dashed line shows the upper limit of the 95\% confidence region. For masses below the cross-over between 
the upper limits  and the theoretical curve, the existence of the hypothesised particle is excluded; for higher 
masses, the data does not distinguish between the two hypotheses $H_1$ and $H_0$.
(d) The 95\% CL region is below the predicted cross-section at all masses, and so the hypothesised new particle is excluded.   
(e) The upper edge of the 95\% CL region lies above the predicted cross-section at all masses, and so the
data does not distinguish between existence or not of the new particle.
\label{fig:my_plots}}
\end{center}
\end{figure}

%

%
\end{document}